\documentclass{jfm}
\usepackage{graphicx}
\usepackage{epstopdf, epsfig}
\usepackage{nicefrac}
\usepackage{amsmath}
\usepackage{savesym}
\savesymbol{nomenclature}
\usepackage{nomencl}
\restoresymbol{OWN}{nomenclature}

\newcommand{\bVec}[1]{\mathbi{#1}}
\newcommand{\Eq}[1]{Eq. (\ref{eq:#1})}
\newcommand{\eq}[1]{Eq. (\ref{eq:#1})}

\newcommand{\figwidth}[0]{0.75\linewidth}

\RequirePackage{ifthen}
   \renewcommand{\nomgroup}[1]{%
     \ifthenelse{\equal{#1}{G}}{}{%
     	\ifthenelse{\equal{#1}{L}}{}{%
         \ifthenelse{\equal{#1}{S}}{\item[]\item[\textit{Sub/superscripts}]}{}}}}

\shorttitle{Wall boundary conditions for direct precipitation fouling modelling}
\shortauthor{S. G. Johnsen, T. M. P\"{a}\"{a}kk\"{o}nen, S. Andersson, S. T. Johansen and B. Wittgens}

\title{On the wall boundary conditions for species-specific mass conservation equations in mathematical modelling of direct precipitation fouling from supersaturated, multi-component fluid mixtures}

\author{Sverre G. Johnsen\aff{1}
  \corresp{\email{sverre.g.johnsen@sintef.no}},
  Tiina M. P\"{a}\"{a}kk\"{o}nen\aff{2},
  Stefan Andersson\aff{1},
  Stein Tore Johansen\aff{1},
 \and Bernd Wittgens\aff{1}}

\affiliation{
  \aff{1}SINTEF Materials and Chemistry, Trondheim, Norway
  \aff{2}University of Oulu, Faculty of Technology, Environmental and Chemical Engineering, Oulu, Finland
}

\begin{document}

\maketitle

\begin{abstract}
The mathematical modelling of species transport in the turbulent boundary layer of fluids that precipitate on the wall, is an important topic at the heart of one of the biggest challenges in efficient energy utilization in all process industries; namely the fouling of heat exchangers. 
A major contributor to the complexity of the problem is the multi-length-scale nature of the governing phenomena. 
That is, transport mechanisms dominating at the nano-scale may be responsible for the macroscopic performance of the industrial process. 
This paper addresses some of the challenges that need to be met in modelling the boundary conditions, i.e. the atomic/molecular-scale conditions, for the species-specific mass conservation equations, at the wall, for single-phase, multi-component fluids that precipitate at the wall.
\end{abstract}

\begin{keywords}
Authors should not enter keywords on the manuscript, as these must be chosen by the author during the online submission process and will then be added during the typesetting process (see http://journals.cambridge.org/data/\linebreak[3]relatedlink/jfm-\linebreak[3]keywords.pdf for the full list)
\end{keywords}

\section{Introduction}\label{sec:introduction}
Fouling of solid surfaces and heat exchanger surfaces in particular, is a common and much studied problem in most process industries \citep{MullerSteinhagen11}. 
Fouling is defined as the unwanted accumulation of solid (or semi-solid) material on solid surfaces. 
Similar is the desired accumulation of solids e.g. in chemical vapour deposition (CVD) \citep{Kleijn89,Krishnan94}. 
A common and costly problem in many industrial applications is the direct precipitation of super saturated fluids on heat exchanger surfaces. 
Typical examples are found in e.g. waste incineration, metal production, or in power plants, where efficient heat recovery is key to sustainable production, and a combination of direct precipitation and deposition of e.g. solid metal oxides is a major show-stopper. 
By precipitation, we understand all types of phase transitions from a fluid to a relatively denser phase, e.g., gas $\to$ liquid (condensation), gas $\to$ solid (sublimation), liquid $\to$ solid (solidification). Our work is relevant for both physisorption and chemisorption.

In previous papers, we have presented the mathematical modelling of solid deposition processes \citep{Johansen91,Johnsen09,Johnsen15}. 
The most recent work \citep{Johnsen15} has been on developing a generic modelling framework for the dissolved species mass transport through the turbulent, reactive boundary layer of multi-component fluid mixtures that precipitate on the wall. 
The modelling is based on Maxwell-Stefan diffusion in multi-component mixtures, and the turbulent, single-phase Navier-Stokes equations. 
The governing equations are simplified in accordance with common assumptions of computational fluid dynamics (CFD), and the developed framework can be employed as a sub-grid model for direct precipitation/ crystallization/solidification fouling in coarse grid CFD models.

\begin{figure}
  \centerline{\includegraphics[width=\figwidth]{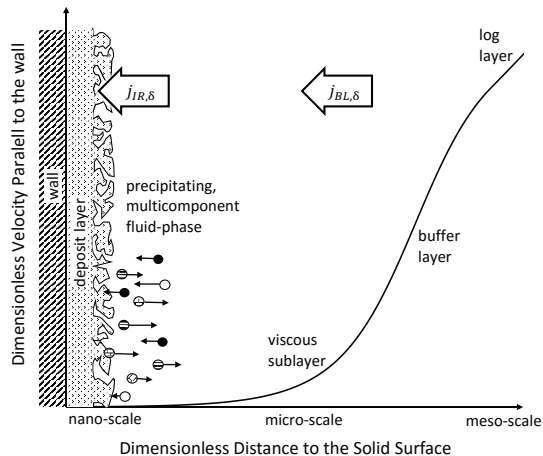}}
  \caption{
    Illustration to demonstrating the various length scales applying in fouling, in the context of the "law of the wall", for the velocity parallel to the wall. 
    The over-all mass flux, $j_{BL,\delta}$, represents the transport of a depositing species, $\delta$, through the turbulent boundary layer, and the interface reaction rate, $j_{IR,\delta}$, represents the mass  integration flux of the species, $\delta$, into the deposit layer at the wall.
  }
\label{fig:Fig1}
\end{figure}

The deposition rate is determined by transportation processes taking place at three main length scales (see figure \ref{fig:Fig1}): 
1) The nano-scale determines if deposits will stick to the wall as well as the rate of phase-change at the wall. 
2) The micro-scale determines the efficiency of turbulent/diffusive transport through the laminar sub layer close to the wall. 
3) The meso-scale determines the bulk conditions at the outskirts of the turbulent boundary layer and transport in the log-layer. 
Our work has so far focused on the micro- and meso-scales. This paper discusses some of the aspects of the processes taking place at the nano-scale, and argues that these are crucial for the accurate/predictive modelling of precipitation from industry-scale process streams. 

The main motivation for this paper is to discuss the importance of understanding the detailed precipitation kinetics taking place at the wall, in order to utilize correct boundary conditions for the species-specific mass conservation equations.  
In order to do this, we need to understand how the boundary conditions for the non-depositing species arise, and what the main differences between the non-depositing and depositing species are, with respect to boundary conditions. 
Finally, we suggest a way forward in how to obtain useful boundary conditions for the depositing species. 

Throughout the paper we will employ the notation wall, when discussing the properties at the wall-fluid interface. 
That is, even if a thick deposit layer is present between the actual wall and the flowing fluid, and a more accurate term would be e.g. the \emph{moving} or \emph{time-evolving solid-fluid interface}, we will, for consistency, use the index $w$ and refer to the wall when discussing the boundary condition seen by the fluid, at the solid-fluid interface. 
Thus, it should be kept in mind that the wall properties, as we are referring to them, are, in reality, not constant values, but rather time-dependent and depending on the deposit layer thickness, porosity, density, etc. 
It is out of scope for the current paper to go into all these details, however, and wall properties will be treated as constants without further consideration of the exact physical or chemical nature of the solid-fluid interface. 

\section{Theory}\label{sec:theory}
We are considering a single-phase fluid mixture consisting of a set, $\Omega$, of $N$ unique, distinguishable, chemical species. 
The set of species furthermore consists of two non-intersecting subsets, namely the set of depositing species, $\Delta$, and the set of non-depositing species, $H$, such that $\Omega=\Delta\cup H$, but $\Delta\cap H=\varnothing$. 
We will reserve the indexes $\delta \in \Delta $ and $\eta \in H $ to denote, exclusively, the species in the sets of depositing and non-depositing species, respectively.
It is assumed that each species field, hence the fluid itself, can be modelled as a continuum. 
This implies that species properties can be considered as continuously varying physical fields that are well defined throughout the fluid domain. 
Furthermore, we assume homogeneous mixing in the sense that local species properties are taken as volume averages over infinitesimal volumes. 
These assumptions allow us to utilize differential calculus in deriving governing equations for the species transport.

\subsection{Definitions}\label{sec:definitions}
An arbitrary species, contained in $\Omega$, and denoted by $i\in \left\{ 1,\ldots ,N \right\}$ moves with the local convective velocity $\bVec{u}_i$, and its local convective mass flux is defined as $\bVec{j}_i=\rho_i\bVec{u}_i$, where $\rho_i$ is the species' mass concentration. 
The advective mass flux is given by $\bVec{j}=\sum\nolimits_{i=1}^{N}{\bVec{j}_i}=\rho_f\bVec{u}_f$, where $\rho_f\equiv\sum\nolimits_{i=1}^{N}\rho_i$is the fluid mass density. 
 The mass-averaged advective velocity is defined as $\bVec{u}_f\equiv\sum\nolimits_{i=1}^{N}{X_i\bVec{u}_i}$, where the species mass fractions are expressed as $X_i=\nicefrac{\rho_i}{\rho_f}$ and can be related to the species mole fractions, $z_i$, via $X_i=\left(\nicefrac{M_{w,i}}{\overline{M}_{w}}\right)z_i$, where $M_{w,i}$ are the species molar masses, and the mean molar mass is given by $\overline{M}_w=\sum\nolimits_{i}{M_{w,i}z_i}$. 
The species' diffusive mass flux is given by the difference between the species' convective and advective mass fluxes;
\begin{equation}\label{eq:difffluxdef}
  \bVec{j}_{i,d}=\rho_i\left(\bVec{u}_i-\bVec{u}_{f}\right)~.
\end{equation}
It follows that the diffusive mass fluxes sum to zero.

\subsection{Mass conservation equation}\label{sec:mass_conservation_equation}
The steady-state, mass conservation equation for the species $i$ is the Advection-Diffusion Equation,
\begin{equation}\label{eq:ADE}
	\bVec{\nabla}\boldsymbol\cdot\left(\rho_i\bVec{u}_{i}\right) =  \bVec{\nabla}\boldsymbol\cdot\left(\rho_fX_i\bVec{u}_f\right) + \bVec{\nabla}\boldsymbol\cdot\bVec{j}_{i,d} = {{R}_{i}}~,
\end{equation}
which, by summation over the species index, $i$, becomes the continuity equation for the fluid mixture,$\bVec{\nabla}\left(\rho_f\bVec{u}_f\right)=0$. $R_i$ is a source-term due to chemical reactions.

\subsection{Diffusion}\label{sec:diffusion}
The diffusive mass flux of the species $i$ can be modelled by Maxwell-Stefan theory,
\begin{equation}\label{eq:MSdiff}
  \bVec{j}_{i,d}=-\rho_f\sum\nolimits_{j=1}^{N-1}{D_{ij}\bVec{d}_j}~,
\end{equation}
where the $D_{ij}$ are the binary diffusivities of components $i$ in components $j$, and the driving force vector is proportional to the gradient of a scalar potential energy field, namely the species' total chemical potential, $\bVec{d}_j=\left(\nicefrac{1}{\mathcal{R}T}\right)\bVec{\nabla}\mu_{tot,j}$. 
It follows from the definitions above, that the diffusive mass flux of the $N$th, dependent species is given by $\bVec{j}_{N,d}=-\sum\nolimits_{i=1}^{N-1}{\bVec{j}_{i,d}}$. 
The dependent species is conveniently referred to as the solvent whereas the remaining species are referred to as the solutes.

We consider, now, a total chemical potential that can be split into a pressure-, temperature- and composition-dependent chemical term, and terms containing contributions from supplementary fields, $\mu_{tot,j} = \mu_{chem,j}\left(P,T,\left\{z_j\right\}\right) + \sum{\mu_{\psi ,j}\left(\psi \right)}$. 
The supplementary potential fields, may be due to externally imposed fields (e.g. gravitational or accelerating motion, or electrical/magnetic fields), but may also be due to internally induced fields (e.g. electric fields in inhomogeneous ionic mixtures). 
Considering, for simplicity, just one supplementary field, we can expand \Eq{MSdiff} in terms of its partial derivatives, and express the diffusive mass flux in terms of field gradients;
\begin{equation}\label{genfluxvec}
  \bVec{j}_{i,d} = -\rho_fD_{ij}\left[d_{T,j}\bVec{\nabla}\left(\ln T\right) + d_{P,j}\bVec{\nabla}P+d_{\psi ,j}\bVec{\nabla}\psi +\Gamma_{jk}\bVec{\nabla}z_k\right]~,
\end{equation}
where $d_{T,j}\equiv\nicefrac{\partial_{T}\mu_{chem,j}}{\mathcal{R}}$, $d_{P,j}\equiv\nicefrac{\partial_{P}\mu_{chem,j}}{\mathcal{R}T}$, $d_{\psi ,j}\equiv\nicefrac{\partial_{\psi }\mu_{\psi ,j}}{\mathcal{R}T}$, and $\Gamma_{jk}\equiv\nicefrac{\partial_{z_k}\mu_{chem,j}}{\mathcal{R}T}$, and Einstein's summation convention is employed. 
We realize that the net diffusive driving force is a combination of a temperature gradient force (thermophoresis), a pressure gradient force (barophoresis), supplementary potential gradient forces (e.g. gravitophoresis, electrophoresis, etc.), and the compositional gradient force (diffusiophoresis). 
Thermophoresis and barophoresis are related to the mixture molar entropy and volume, respectively. 
The diffusiophoretic contribution is related to the species' activities and involves the compositional gradients of all the species in the mixture, but reduces to the Fickian diffusion term under certain conditions (e.g. binary or dilute mixtures, or mixtures where the constituents are similar in size and nature \citep{Taylor93}). 
Other phoretic behaviours are related to the species charges in relation to the supplementary fields (e.g. an ion species of molar electric charge $Q_i$ and  mass ("gravitational charge") $M_{w,i}$, will experience molar electrophoretic and gravitophoretic driving forces, $\left(\nicefrac{Q_i}{RT}\right)\bVec{\nabla}\psi_E$ and $\left(\nicefrac{M_{w,i}}{RT}\right)\bVec{\nabla}\psi_G$, in the electrical and gravitational potential fields, $\psi_E$ and $\psi_G$, respectively). 
Converting the mole fractions to mass fractions, $z_k=\Lambda_{kl}X_l$, where $\Lambda_{kl}=\overline{M}_w\left[\nicefrac{z_k}{M_{w,N}} + \nicefrac{\left(\delta_{kl}-z_k\right)}{M_{w,l}}\right]$, and $\delta_{kl}$ is the Kronecker delta, the diffusive mass flux of species $i$, in the $\alpha$ direction, is expressed as
\begin{equation}\label{eq:genflux}
  j_{i,d,\alpha} = -\rho_f\left[D_{T,i}\partial_\alpha\left(\ln T\right) + D_{P,i}\partial_\alpha P + D_{\psi ,i}\partial_\alpha\psi + D_{X,il}\partial_\alpha X_l\right]~,
\end{equation}
where $D_{T,i}\equiv D_{ij}d_{T,j}$ , $D_{P,i}\equiv D_{ij}d_{P,j}$, $D_{\psi ,i}\equiv D_{ij}d_{\psi ,j}$, and $D_{X,il}\equiv D_{ij}\Gamma_{jk}\Lambda_{kl}$.

\subsection{Advection}\label{sec:advection}
Due to the no-slip condition, effects of turbulence vanish close to the wall, and molecular diffusion is the only remaining transport mechanism. 
To conserve the total mass of the mixture, however, a non-zero, advective velocity component normal to the wall is required, as will be shown below. 
In the following, we will consider mass fluxes in the direction normal to the wall ($\bot$), only. The total deposition flux is found by summing \eq{difffluxdef}, evaluated at the wall, over all species,
\begin{equation}\label{eq:jdep}
  j_{dep}\equiv \rho_{f,w}u_{f,\bot ,w}=\sum\nolimits_\delta{\rho_{\delta ,w}u_{\delta ,\bot ,w}}~,
\end{equation}
 and using that 1) the convective mass fluxes of the non-depositing species are zero at the wall; and 2) that the diffusive mass fluxes sum to zero. 
It follows directly that the advective fluid velocity at the wall is given by the sum of the mass fraction weighted convective velocities of the depositing species,
\begin{equation}\label{eq:advvel}
  u_{f,\bot ,w}=\sum\nolimits_\delta{X_{\delta ,w}u_{\delta ,\bot ,w}}~,
\end{equation}
and employing the definition in \eq{difffluxdef}, it follows that \eq{advvel} can be expressed as
\begin{equation}\label{eq:advvelfrac}
  u_{f,\bot ,w}=\frac{\sum\nolimits_\delta{j_{\delta ,d,\bot}}}{\rho_{f,w}\left(1 - \sum\nolimits_\delta{X_{\delta ,w}}\right)}~.
\end{equation}
Thus, the advective fluid velocity at the wall is uniquely defined by the diffusive mass fluxes and mass fractions of the depositing species, at the wall.

\subsection{Species mass fraction boundary conditions at the wall}\label{sec:species_mass_frac_bc}
Provided the convective mass fluxes of the non-depositing species vanish at the wall, \eq{difffluxdef} dictates that the diffusive mass fluxes of the non-depositing species be exactly cancelled by the advective mass flux, at the wall;
\begin{equation}\label{eq:fluxnondep}
  j_{\eta ,d,\bot ,w}=-\rho_{\eta ,w}u_{f,\bot ,w}~.
\end{equation}
By inserting for equations \ref{eq:genflux} and \ref{eq:advvel} in \eq{fluxnondep}, relationships between the mass fraction gradient of the non-depositing species, the mass fractions of the depositing species, and the temperature, pressure, and supplementary field gradients, evaluated at the wall, are obtained; 
\begin{equation}\label{eq:nondepBC}
  \left[-D_{T,\eta }\partial_\bot\left(\ln T\right) - D_{P,\eta}\partial_\bot P - D_{\psi ,\eta }\partial_\bot\psi - D_{X,\eta l}\partial_\bot X_l + X_\eta\sum\nolimits_\delta{X_\delta u_{\delta ,\bot}}\right]_w=0~.
\end{equation}
\Eq{nondepBC} acts as the wall boundary condition for \eq{ADE}, for the arbitrary, non-depositing species $\eta$. 
It provides one equation for the mass fraction gradients for each of the non-depositing solute species. 
The solution of this linear system of equations provides the Neumann boundary conditions, for the mass conservation of the non-depositing species, at the wall. 
We realize, however, that \eq{nondepBC} needs estimates of the temperature, pressure and supplementary potential gradients, the various diffusivities of all the species, the mass fractions of all the depositing species, and finally the convective velocities of the depositing species, at the wall. 
Thus, these will have to be updated for each iteration when solving the governing equations numerically. 
Their profiles are, in fact, part of the final solution to the problem.
The possibility to derive \eq{nondepBC} for the non-depositing species arises from the fact that we know their mass fluxes at the wall. 
The mass fluxes of the depositing species, however, are a priori unknown. 
Thus, there is no such simple approach to find their boundary conditions. 
The natural choice is to impose Dirichlet boundary conditions on the depositing species. 
In the Discussion section, below, we will make some comments on how to obtain these numbers.

\subsection{One single depositing species}\label{sec:one_single_dep_sp}
In the special case that only one of the species deposits, the summation signs in equations \ref{eq:jdep} and \ref{eq:advvel} are superfluous, and we can express the total deposition flux as 
\begin{equation}
  j_{dep} \equiv \rho_{f,w}u_{f,\bot ,w}=\rho_{\delta ,w}u_{\delta ,\bot ,w}~.
\end{equation}
\Eq{advvelfrac} can be expressed in terms of the diffusive mass flux of the depositing species and the difference between the fluid mass density and the depositing species' mass concentration, at the wall; 
\begin{equation}\label{eq:advveldeprate}
  u_{f,\bot ,w}=\frac{j_{\delta ,d,\bot ,w}}{\rho_{f,w}-\rho_{\delta ,w}}~.
\end{equation}
Inserting for \eq{advveldeprate} in \eq{fluxnondep}, we get a relationship between the non-depositing and depositing species' diffusive mass fluxes at the wall;
\begin{equation}\label{eq:jnondepjdep}
  j_{\eta ,d,\bot ,w}=-\frac{X_{\eta ,w}}{1-X_{\delta ,w}}j_{\delta ,d,\bot ,w}~.
\end{equation}
Now, in the case of one single depositing species, the combination of equations \ref{eq:genflux} and \ref{eq:jnondepjdep} reveals that we must require that the temperature, pressure, supplementary potential, and mass fraction gradients at the wall, balance in accordance with the equation
\begin{multline}\label{eq:singledepspbc}
  \Bigg[\left(1 + \frac{X_{\eta }}{1-X_{\delta }}\frac{D_{T,\delta }}{D_{T,\eta }}\right)D_{T,\eta }\partial_{\bot }\left(\ln T\right) + \left(1 + \frac{X_{\eta }}{1-X_{\delta }}\frac{D_{P,\delta }}{D_{P,\eta }}\right)D_{P,\eta }\partial_{\bot }P \\ 
    + \left(1 + \frac{X_{\eta }}{1-X_{\delta }}\frac{D_{\psi ,\delta }}{D_{\psi ,\eta }}\right)D_{\psi ,\eta }\partial_{\bot }\psi + \sum\limits_{\begin{smallmatrix} 
  l=1, \\ 
  l\ne \delta  
 \end{smallmatrix}}^{N-1}{\left(1 + \frac{X_{\eta }}{1-X_{\delta }}\frac{D_{X,\delta l}}{D_{X,\eta l}}\right)D_{X,\eta l}\partial_{\bot }X_{l}}\Bigg]_{w}=0~,
\end{multline}
 for each non-depositing species, $\eta$. In the case of one single depositing species, \eq{singledepspbc} replaces \eq{nondepBC} in determining the Neumann boundary conditions for the non-depositing species. 
Furthermore, in the special case that the fluid consists of two species, only, where one is depositing, and the other not, we can express $X_{\eta }=1-X_{\delta }$, and \eq{singledepspbc} reduces to
\begin{equation}
  \left[\partial_T\left(\mu_{chem,\eta } +\mu_{chem,\delta }\right)\partial_{\bot }T + \partial_P\left(\mu_{chem,\eta} + \mu_{chem,\delta }\right)\partial_{\bot}P + \partial_\psi\left(\mu_{\psi,\eta} + \mu_{\psi,\delta}\right)\partial_{\bot}\psi\right]_w=0~.
\end{equation}

\subsection{Simplified boundary conditions for the non-depositing species}\label{sec:simplified_bc}
In the case that the advective fluid velocity at the wall is negligible, the boundary conditions simplify significantly. 
In this case, it suffices to require that the wall-normal diffusive mass fluxes of the non-depositing species vanish, at the wall. 
Thus, we require that the wall-normal right-hand-side terms in \eq{genflux} cancel out, and we get the simplified Neumann wall boundary condition for the non-depositing species;
\begin{equation}
  \left.\partial_\bot z_\eta\right|_w = -\left[\left(\Gamma_{i\eta}\right)^{-1}\left(d_{T,i}\partial_\bot\ln T + d_{P,i}\partial_\bot P + d_{\psi ,i}\partial_\bot\psi\right)\right]_w~. 
\end{equation}

\section{Discussion}\label{sec:discussion}
From the theoretical considerations above, two questions naturally arise:
\begin{enumerate}
   \item How can we estimate the mole fractions of the depositing species, at the wall?
   \item How do we know which species are depositing and which are non-depositing, in the first place?
\end{enumerate}
These questions will only partly be answered in this paper, but a discussion of what is required to handle these questions is provided below. 
To be able to answer these questions, a profound understanding of the fundamentals of the precipitation kinetics at a molecular/atomic level, at the wall and at the deposit layer surface, is required. 
This understanding can only be obtained through the combination of experimental studies and atomic-scale modelling of the solid-fluid interface reactions (e.g. atomistic/molecular dynamics simulations). 
Then, it is required to translate modelling results from molecular dynamics calculations into the language of thermodynamics, to be able to utilize the chemical potential gradient in the calculation of the deposition fluxes. In particular, the partial derivative of the chemical potential with respect to temperature, or rather the Soret coefficient responsible for the thermophoretic behaviour, is currently poorly understood.

The direct precipitation fouling process consists of two major steps, namely the mass transport to the wall and the phase change (a.k.a. surface integration) at the wall. 
The former is governed by diffusive, advective, and turbulent transport mechanisms, in the fluid phase. 
The latter depends on the thermodynamic/chemical integration of fluid phase species into the foulant layer. 
In addition, other processes may take place simultaneously; diffusion and counter diffusion in the porous foulant layer; adhesion probability less than one due to short residence times at the wall; and re-entrainment of deposited material \citep{Bott95}. 
If the time-scale of mass transport to the wall is much longer than that of the phase-change, the fouling process is characterized as diffusion controlled. 
The regime is characterized by e.g. increasing fouling rate with increasing flow velocity. 
Opposite, if the phase-change at the wall is limiting the fouling rate, the fouling process is denoted interface controlled. 
The regime is characterized by constant (or even decreasing) fouling rate with increasing flow velocity. 
In the literature there are reports of both diffusion controlled \citep{Hasson68} and interface controlled \citep{Augustin95,Bansal93,Mwaba06a,Mwaba06b} fouling regimes. 

Direct precipitation fouling has traditionally been modelled in accordance with the Kern-Seaton approach, where the two main items for deposition, are the boundary layer mass transfer and the surface integration. 
These are typically modelled by two rate equations, for the depositing species, $\delta$, where the mass transfer rate is proportional to some power of a concentration difference \citep{MullerSteinhagen11, Bott95, Konak74}. 
For the boundary layer:
\begin{equation}\label{eq:jBL}
  j_{BL,\delta}=k_{BL,\delta}\left(X_{\delta ,Bulk}-X_{\delta ,w}\right)~,
\end{equation}
where $k_{BL}$is an effective mass transfer coefficient due to all transport phenomena in the boundary layer, and $X_{\delta ,Bulk}$ and $X_{\delta ,w}$ are the bulk and solid-fluid interface (wall) mass fractions of species $\delta$, respectively. 
For the surface integration:
\begin{equation}\label{eq:jIR}
  j_{IR,\delta}=k_{IR,\delta}\left(X_{\delta ,w}-X_{\delta ,Sat,w}\right)^{n_{IR,\delta}}~,
\end{equation}
where $k_{IR}$ is an effective mass transfer coefficient at the solid-fluid interface (interface reaction rate) and $X_{\delta ,Sat,w}$ is the saturation mass fraction of species $\delta$, at the solid-fluid interface. 
Mass conservation requires that $j_{BL}=j_{IR}$, which determines the interface mass fraction. 
E.g., by linearising \eq{jIR}, it can be shown that the wall mass fractions of species $\delta$ can be approximated by
\begin{equation} 
  X_{\delta ,w}\approx\frac{k_{BL,\delta }X_{\delta ,bulk}-A_{IR,\delta}}{k_{BL,\delta}+B_{IR,\delta}}~,
\end{equation}
resulting in a mass flux of
\begin{equation}
  j_{\delta}=j_{BL,\delta }=j_{IR,\delta }\approx\frac{k_{BL,\delta}\left(B_{IR,\delta}X_{\delta,bulk}+A_{IR,\delta}\right)}{k_{BL,\delta}+B_{IR,\delta}}~,
\end{equation}
where $A_{IR,\delta}\equiv\frac{B_{IR,\delta}}{n_{IR,\delta}}\left[\left(1-n_{IR,\delta}\right)X_{\delta,0}-X_{\delta,Sat,w}\right]$, $B_{IR,\delta}\equiv n_{IR,\delta}k_{IR,\delta}\left(X_{\delta,0}-X_{\delta,Sat,w}\right)^{n_{IR,\delta}-1}$, and $X_{\delta,0}$ is the approximate wall mass fraction. 
Ultimately, it is the magnitude of the interface reaction rate, $k_{IR,\delta}$, that determines the fouling regime classification. 
That is, in the extremes of fully diffusion- or interface-controlled fouling we have $k_{IR}\to\infty$ and $k_{IR}\to0$, respectively. 
In the fully diffusion-controlled regime, it is evident that \eq{jIR} dictates $X_{\delta,w}\xrightarrow[k_{IR}\to\infty]{}X_{\delta ,Sat,w}$ , to ensure finite interface mass-transfer. 
In the fully interface-controlled regime, \eq{jBL} requires that $X_{\delta,w}\xrightarrow[k_{IR}\to0]{}X_{\delta ,Bulk}$ \citep{Mwaba06a}. 
This modelling approach is, however, only able to capture deposition due to downhill mass diffusion, i.e. where diffusion is in the direction of decreasing concentration, and it relies on supersaturated bulk conditions, i.e. $X_{\delta,Sat}\le X_{\delta,w}\le X_{\delta,Bulk}$. 
It fails to account for the complex behaviour that can arise from the thorough thermodynamic considerations introduced in the Theory section, above. 
That is, the traditional approach is well suited for simple scenarios (e.g. isothermal/isobaric conditions without supplementary fields), but it cannot be expected to produce accurate results in the presence of e.g. strong temperature gradients. 

\begin{figure}
  \centerline{\includegraphics[width=\figwidth]{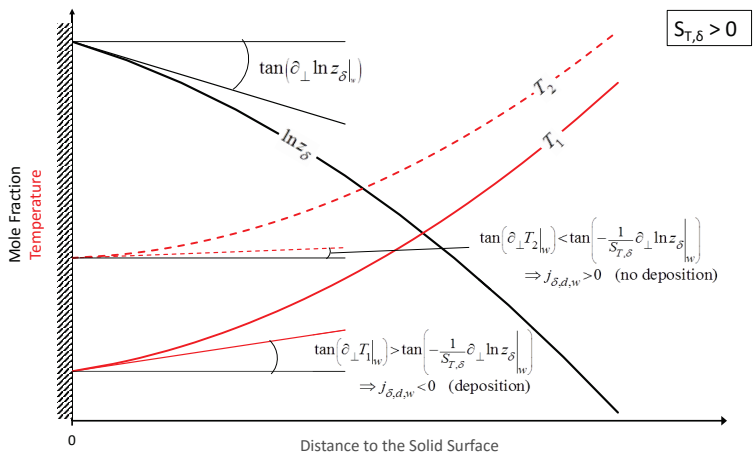}}
  \caption{
    Sketch of two scenarios with a given mole fraction profile decreasing away from the wall (negative gradient), positive Soret coefficient, and two different temperature profiles resulting in negative (deposition) and positive (no deposition) mass fluxes, respectively (See \Eq{StdT}). 
  }
\label{fig:Fig2}
\end{figure}

Even for simple mixtures, however, it is possible to achieve deposition due to uphill diffusion, i.e. where the net diffusive mass flux is in the direction of increasing concentrations of the depositing species. 
E.g., consider a two-component, ideal mixture, where one species can deposit on the wall, under isobaric conditions and the absence of supplementary fields. 
The wall-normal component of \eq{genflux}, for the depositing species, can be written
\begin{equation}\label{eq:jSoret}
  j_{\delta,d,\bot,w}=-\left.\rho_fD_\delta\left[\partial_\bot z_\delta + z_\delta\left(1 - z_\delta\right)S_{T,\delta}\partial_\bot T\right]\right|_w<0~,
\end{equation}
where the overall diffusivity is expressed as $D_\delta\equiv 2\frac{D_{\delta \delta }\left(1 - z_\delta\right) + D_{\delta\eta}z_\delta}{z_\delta\left(1 - z_\delta\right)}$, and the Soret coefficient is given by $S_{T,\delta}=\frac{1}{2RT}\frac{D_{\delta\delta}\partial_T\mu_\delta + D_{\delta\eta}\partial_T\mu_\eta}{D_{\delta\delta}\left(1 - z_\delta\right) + D_{\delta\eta}z_\delta}$ , and the positive mass flux points in the direction away from the wall, into the fluid, per definition. 
In the limit of dilute mixture ($z_\delta\ll1$), and similar diffusivities ($D_{\delta\delta}\approx D_{\delta\eta}$), \eq{jSoret} yields the requirement
\begin{equation}
  \left.S_{T,\delta}\partial_\bot T\right|_w>-\left.\partial_\bot\ln{z_\delta}\right|_w~.
\end{equation}
Now, for a negative composition gradient (mole fraction decreasing in the direction pointing away from the wall), we see that
\begin{equation}\label{eq:StdT}
  \left.S_{T,\delta}\partial_\bot T\right|_w>-\left.\partial_\bot\ln{z_\delta}\right|_w>0~.
\end{equation}
That is, for a steep enough temperature gradient (sign depending on the sign of the Soret coefficient), a positive deposition rate (negative mass flux at the wall) can occur even for negative mole fraction gradients (uphill diffusion). 
This is illustrated in figure \ref{fig:Fig2}, which shows a monotonously decreasing mole fraction profile, $\ln{z_\delta}$, along with two different monotonously increasing temperature profiles, $T_1$ and $T_2$. 
As indicated in the figure, $T_1$is steep enough to give a negative mass flux, i.e. deposition at the wall, for a certain Soret coefficient, $S_{T,\delta}>0$. 
$T_2$, however, results in a positive mass flux, hence no deposition. 
It can be deduced that there exists a critical temperature gradient, $\left.\partial_\bot T_2\right|_w<\left.\partial_\bot T_c\right|_w<\left.\partial_\bot T_1\right|_w$, that results in zero mass flux, $j_{\delta,d,w}=0$. 
In the presence of additional fluid components or supplementary influencing fields, there are virtually endless opportunities for uphill diffusion-based deposition. 
In conclusion, the true upper bound for the solid-fluid interface mass fraction, in a single-phase system, is the critical supersaturation, $X_{\delta,CSat}$, not the bulk mass fraction. 
Even saturations above the critical supersaturation may be permitted if transport equations for additional (solid) phases, dispersed in the fluid phase, are taken into account.

Since the $X_{i,Sat}$ generally are functions of the overall composition, temperature, pressure, and the effect of supplementary, influencing fields, we deduce that $X_{\delta,w}=X_{\delta,Sat,w}$ implies that $X_{i,w}=X_{i,Sat,w}\forall i\in\Omega $. 
This situation is commonly referred to as a situation of local thermodynamic equilibrium at the solid-fluid interface. 
It further implies that the chemical potentials of the solid and fluid are identical at the interface. 
Assuming that the chemical potentials of the depositing species increase with their mass fractions, it will be energetically favourable to undergo a fluid-solid phase-change, and thus deposit on the wall, in the situation where $X_{\delta ,w}>X_{\delta,Sat,w}$. 
In the opposite situation, where $X_{\delta,w}<X_{\delta,Sat,w}$, however, it will be energetically unfavourable. 
In this sense, we accept $X_{\delta,Sat,w}$ as the lower bound for the interface mass fraction. 
Thus, we have established that $X_{\delta,Sat,w}\le X_{\delta,w}\le X_{\delta,CSat}~\forall~\delta\in\Delta$ in single-phase flow, but $X_{\delta,Sat,w}\le X_{\delta,w}\le 1~\forall~\delta\in\Delta$ in a multi-phase framework. 

We are now ready to address the two questions posed at the beginning of the Discussion section. 
First, having established the lower and upper bounds for the depositing species' mass fractions at the solid-fluid interface, we still need to estimate the actual interface mass fractions to be able to solve the governing equations for mass-transport through the boundary layer. 
The interface mass fractions are, however, part of the solution, and thus the estimation of the interface mass fractions must be part of the iterative, numerical scheme to solve the governing equations. 
Employing \eq{jIR}, we can calculate the interface mass fractions from the equation
\begin{equation}
  X_{\delta,w}=X_{\delta,Sat,w} + \left(\nicefrac{j_{\delta,\bot,w}}{k_{IR}}\right)^{\nicefrac{1}{n_{IR}}}~,
\end{equation}
where $X_{\delta,Sat,w}$ must be determined from thermodynamic/chemical equilibrium calculations at wall conditions or from interpolation in tables of experimental data, $j_{\delta,\bot,w}$ is the deposition rate of species $\delta$, from the previous iteration, and the interface reaction coefficient can be determined from e.g. the Arrhenius equation, $k_{IR}=k_{IR,0}\exp\left(\nicefrac{-E_a}{RT_w}\right)$, where the kinetic parameters $k_{IR,0}$ and $E_a$ must be determined experimentally or from molecular dynamics simulations.

To answer which species will deposit, it is necessary to perform a full thermodynamic/chemical equilibrium analysis of the fluid-substrate system, to provide information on e.g. the supersaturation degrees with respect to different solid stages and fluid compositions. 
This is beyond the scope of this paper, but we point the reader's attention in the direction of recent developments in atomic/molecular-scale modelling and simulation. 
There has been done a substantial amount of work on studying different types of atomic and molecular deposition at solid surfaces using molecular dynamics and other atomic-scale modelling techniques 
Results that can be derived from these simulations are, in a rough order of required computational complexity, e.g., nano-scale surface structures, growth and reaction mechanisms, thermodynamics data (such as reaction enthalpies, free energy differences and equilibrium constants), and growth kinetics parameters (rate coefficients and corresponding Arrhenius parameters). 
That is, molecular/atomic scale simulations may provide essential information on how surface structures may affect the formation of solid phases at the surface.

An additional topic that may be solved by employing advanced molecular dynamics simulations is the accurate, predictive modelling of the thermophoretic driving force coefficients (e.g. the Soret coefficients). 
There has in recent years been much progress in applying molecular dynamics simulations for the study of thermophoretic effects \citep{Artola07,Galliero08,Reith00}, including the calculation of Soret coefficients, so the practical application of such methods for quantitative predictions is certainly plausible.  
This is a crucial achievement in the predictive modelling of molecular transport in the turbulent boundary layer, yet it is currently poorly understood, especially for particles and large molecules, but also for smaller molecules. 
E.g., for colloidal particles suspended in a liquid, there are still no models that can even predict the sign of the Soret coefficient \citep{Geelhoed14}. 
The sign of the Soret coefficient is generally a function of temperature, thus, the fluid mixture's deposition behaviour can be very sensitive to the temperature. 
That is, there may exist a critical wall temperature, where the Soret coefficient changes sign, with the implication that deposition can be turned on/off by selecting the wall temperature carefully.

\section{Conclusion}\label{sec:conclusion}
The direct precipitation fouling rate of a supersaturated, multi-component fluid mixture is depending on the flow conditions in the fluid and a host of fluid properties and parameters. 
What ultimately determines the fouling rate, however, is the surface reaction/phase change that takes place at the nano-scale, at the wall, initially, and at the deposit layer surface, eventually. 
To obtain accurate/predictive mathematical models for the direct precipitation fouling rate, it is crucial to establish fundamental models for the solid-fluid interface conditions that act as wall boundary conditions in the species mass conservation equations. 
In this way, we can combine the three very different length-scales at which the transportation of depositing species takes place. 
Thus, even for industrial processes, taking place at very large length scales, it is crucial to understand the atomistic/molecular scale-phenomena that occur at the wall, to fully appreciate predictive fouling models coupled with CFD models.

\printnomenclature

\section*{Acknowledgements}
This work was funded by the Research Council of Norway and The Norwegian Ferroalloy Producers Research Association, through the SCORe project\citep{SCORE}. 

\bibliographystyle{jfm}
\bibliography{References}

\end{document}